\documentclass[letter]{jpsj2} 

\def\pade{ Pad\'e } \def\Omicron{ {\rm O} }  \def\d#1{\,{\rm d}#1} \def\i{{\bf i } }

\title{Theory of Fano-Kondo effect of transport properties through quantum dots}
\author{Isao \textsc{Maruyama}$^{1}$, Naokazu \textsc{Shibata}$^{2}$ and Kazuo \textsc{Ueda}$^{1}$}
\inst{
$^{1}$Institute for Solid State Physics, University of Tokyo,
5-1-5 Kashiwa-no-ha, Kashiwa, Chiba 277-8581, Japan\\
$^{2}$Department of Basic Science, University of Tokyo
3-8-1 Komaba, Meguro, Tokyo 153-8902, Japan\\
}

\kword{quantum dot, Fano effect, Kondo effect, density matrix renormalization group method, maximum entropy method, \pade approximation}
\abst{ The Fano-Kondo effect in zero-bias conductance is investigated based on a theoretical model for the T-shaped quantum dot.  The conductance as a function of the gate voltage is generally characterized by a Fano asymmetric parameter $q$.
With varying temperature the conductance shows a crossover between the high and low temperature regions compared with the Kondo temperature $T_{\rm K}$: two Fano asymmetric peaks at high temperatures and the Fano-Kondo plateau inside a Fano peak at low temperatures.  Temperature dependence of conductance is calculated numerically by the Finite temperature density matrix renormalization group method (F$T$-DMRG).  }

\date{\today}

\begin{document}

\maketitle

When a localized state is coupled with a continuous spectrum phase relation of the interference is important and this phenomenon is known as the Fano effect\cite{Fano61}.  In many cases, Coulomb interaction in the localized state cannot be neglected and it leads to the interesting many-body problem of the Kondo effect.  However in usual Kondo effects of magnetic impurities the former aspect is trivial or not apparent and thus has not been discussed explicitly.

Recently it has become possible to observe Kondo effects in transport properties through quantum dots (QDs).  In the simplest geometry of a QD attached to source and drain leads, an embedded QD system, conductance reaches the unitarity limit $2e^2/h$ at zero temperature for the single channel case in the Coulomb blockade region with odd number of electrons. Actually the unitarity limit due to the Kondo effect was observed experimentally~\cite{WielFFETK00}.

With the use of nano-scale technology it is possible to fabricate various geometries for QDs.  The Fano effect combined with the Kondo effect for a single-level QD has been proposed theoretically in two geometries: a QD in a closed Aharonov-Bohm interferometer (AB-QD)\cite{HofstetterKS01,BulkaS01} and a side-coupled QD\cite{TrioHCP02,KangCKS01,AligiaP02}. The latter geometry is a QD coupled with a quantum wire like a stub and we call it a T-shaped QD in the present paper.

The line shape of zero-bias conductance as a function of the gate voltage is characterized by the Fano asymmetric parameter $q$. For the embedded QD, the conductance has a symmetric peak structure, which corresponds to infinite $q$.  On the other hand conductance for the T-shaped QD sometimes shows a symmetric dip structure, which means $q=0$. In this case due to the Kondo effect the conductance becomes zero at $T=0$ and the system shows a perfect reflection rather than the resonance tunneling for the infinite $q$.  This phenomenon is called as an anti-Kondo resonance in literature~\cite{KangCKS01}.  For an AB-QD, the Fano parameter takes a non-zero value.  It can be controlled by the magnetic flux penetrating in the AB ring and can be a complex number.  Experimentally, tunability of the Fano parameter in the AB-QD has been demonstrated successfully~\cite{KobayashiAKI02}.
Fano effects with finite $q$ are general phenomena in quantum transport and have been observed in various geometries not only in the AB-QD but also in single electron transistors~\cite{GoresGHK00} and T-shaped QDs~\cite{KobayashiASKI03}.  However, so far, there has been no experimental report on indication of the Kondo effect for the cases with finite $q$ except for the very recent experiment of the anti-Kondo resonance~\cite{FanoQD:Katumoto}.

Concerning theories for the T-shaped QD, the Fano-Kondo effect studied in ref. \citen{TrioHCP02,KangCKS01,AligiaP02} has been limited to the $q=0$ case.  In this letter we study the Fano-Kondo effect covering the whole range of $q$ by considering an extended theoretical model for the T-shaped QD.  In particular temperature dependence of the conductance is discussed for finite $q$ and finite $U$.  We note that the theories on the Fano-Kondo effect for the AB-QD have been limited to either the $T=0$ case~\cite{HofstetterKS01} or the $U=\infty$ case~\cite{BulkaS01}.  
It is also shown in this letter that the Fano-Kondo effect of the AB-QD can be described in the same way by the mapping to the extended model.

\begin{figure}[tb]
\resizebox{8cm}{!}{\includegraphics{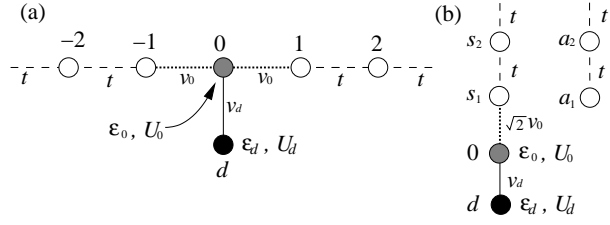}}
\caption{(a) The model for the T-shaped quantum dot. (b) Mapping to two semi-infinite chains of the symmetric $s_i$ and antisymmetric $a_i$ sites, $s_i=(c_i+c_{-i})/\sqrt{2}$ and $a_i=(c_i-c_{-i})/\sqrt{2}$.  }
\label{f-1}
\end{figure}

We introduce the model for the T-shaped QD shown in Fig.~\ref{f-1} (a).  Since we have introduced the energy level $\epsilon_0$ for the $0$-th site the system may be considered as a T-shaped double dot system~\cite{KimH01,KikoinA01}.  The Hamiltonian of the model is
\begin{eqnarray}
{\cal H}&=& - \sum_{i\sigma} t_{i,i+1} ( c^\dagger_{i\sigma} c_{i+1\sigma} + \mbox{h.c.} ) \nonumber \\ && +\epsilon_0 \sum_{\sigma} c^\dagger_{{0}\sigma} c_{{0}\sigma} +U_0 c^\dagger_{0\uparrow} c_{0\uparrow} c^\dagger_{0\downarrow} c_{0\downarrow} \nonumber \\ && -v_d \sum_{\sigma} ( d^\dagger_{\sigma} c_{0\sigma} + \mbox{h.c.} ) \nonumber \\ && +\epsilon_d \sum_{\sigma} d^\dagger_{\sigma} d_{\sigma} +U_d d^\dagger_{\uparrow} d_{\uparrow} d^\dagger_{\downarrow} d_{\downarrow},
\label{eq:H}
\\ t_{i,i+1}&:=& \left\{
\begin{array}{cc}
v_0 &\mbox{ ,when $i=0$ or $-1$, } \\ t &\mbox{ ,others.}, \\
\end{array}
\right.
\end{eqnarray}
where each dot has a single level, $\epsilon_0$ and $\epsilon_d$, and the on-site Coulomb interaction, $U_0$ and $U_d$.  Normalized conductance $g=G \left/ \left ({2e^2\over h}\right)\right.$ is given by
\begin{eqnarray}
g&=& \int^{\infty}_{-\infty} T (\omega) \left (-{\partial f (\omega) \over \partial \omega}\right) {\rm d}\omega
\label{eq:G}
\label{eq:g}
\\ &=& \int^{\infty}_{-\infty} \Delta_0 (\omega) (-\mbox{Im}[G_{0}(\omega)])\left (-{\partial f (\omega) \over \partial \omega}\right) {\rm d}\omega,
\end{eqnarray}
where $\Delta_0 (\omega)= \Delta_0 \sqrt{1 -\left ({\omega \over 2t}\right)^2 }$ with $\Delta_0= 2 v_0^2/t$ and $G_0 (\omega)$ is the local Green's function of the $0$-th site.  These formulae give conductance for the embedded QD when we cut the connection to the $d$-dot by putting $v_d=0$.  In the following, we will concentrate on the T-shaped QD and $U_0=0$ is assumed unless otherwise stated.  We also assume that the chemical potential is zero, which is the half-filling condition because the energy band of the bulk is $-2t\cos k$.

For the T-shaped QD model we can rewrite the conductance by $G_{d}$, because $G_{d}$ and $G_0$ are related by
\begin{eqnarray}
G_0 (\omega)&=&g_0 (\omega)\left (1+v_d^2 g_0 (\omega) G_d (\omega) \right).
\label{eq:Grel}
\end{eqnarray}
The Green's function $g_0$ is the local Green's function at the site $0$ for the case of $v_d=0$ and is given by
\begin{eqnarray}
g_0 (\omega)&=& \left (\omega-\epsilon_0 - 2v_0^2 g_{\rm B}(\omega) \right)^{-1},
\end{eqnarray}
where $g_{\rm B}$ is the local Green's function at the $i=\pm 1$ site for the case of $v_0 =0$, i.e. at the end point of the semi-infinite chain:
\begin{eqnarray}
g_{\rm B}(\omega)&=& \int_{-2t}^{2t} \d{z} { (2\pi t^2)^{-1}\sqrt{4t^2 - z^2} \over \omega-z+\i 0_+}.
\end{eqnarray}
Since we can expand $g_0 (\omega)$ as $g_0 (\omega)\simeq g_0 (0) + g_0 (0)^2 (1+ \Delta_0/t) \omega + \Omicron (\omega^2)$, we can use $g_0 (\omega)=g_0 (0)$ at low temperatures, $T\ll \Delta_0$.  With using the limiting value $g_0 (\omega)=g_0 (0)=-(\epsilon_0-i\Delta_0)^{-1}$, the transmission probability at $T\ll \Delta_0$ and $t$ is written as
\begin{eqnarray}
T (\omega)&=&{1\over 1+q^2} - {2q \over 1+q^2}\Delta_d \mbox{Re}[ G_{d}(\omega)] \nonumber \\&& - { q^2 -1 \over 1+ q^2} \Delta_d \mbox{Im}[ G_{d}(\omega)],
\label{eq:TlowT}
\end{eqnarray}
where the Fano parameter $q$ is
\begin{eqnarray}
q&=& {\epsilon_0 \over \Delta_0},
\end{eqnarray}
and the width of the $d$-level is given by
\begin{eqnarray}
\Delta_d&=& { v_d^2 \Delta_0 \over \epsilon_0^2 + \Delta_0^2}.
\label{eq:non-int:Delta_d}
\end{eqnarray}

It should be noted that in the present model the conductance has a contribution from the real part of $G_d (\omega)$ for non-zero $q$.  Especially at $|q|=1$, only $\mbox{Re}[G_d (\omega)]$ contributes to the conductance.  For other two limiting cases, $q=0$ and $|q|\simeq \infty$, the conductance is determined solely by $\mbox{Im}[G_d (\omega)]$.  The former case, $q=0$, is called anti-resonance case where the transmission probability is given by
\begin{eqnarray}
T (\omega)&=&1-\left (-\Delta_d \mbox{Im} G_d (\omega) \right).
\label{eq:q0}
\end{eqnarray}
The latter case, $|q|\simeq \infty$, is written as
\begin{eqnarray}
T (\omega)&\simeq&-\Delta_d \mbox{Im} G_d (\omega).
\end{eqnarray}
It is worth to mention that conductance in the embedded QD corresponds to the $|q|= \infty$ case,
\begin{eqnarray}
T (\omega)&=&-\Delta_0 \mbox{Im} G_0 (\omega).
\label{eq:embedded}
\end{eqnarray}
Equivalence between the embedded QD and the $|q|=\infty$ case of the T-shaped QD may be understood in the following way.  In the limit of $|q|\rightarrow \infty$, we can write $\Delta_d={2\over t}\left({v_0 v_d \over \epsilon_0}\right)^2$, which shows the second order process plays the role of effective hopping between the $d$-dot and the leads.

It is instructive to consider a mapping from the Hamiltonian of the T-shaped QD to two semi-infinite chains.  The mapping is obtained by considering the symmetric $s_i=(c_i+c_{-i})/\sqrt{2}$ and the antisymmetric $a_i=(c_i-c_{-i})/\sqrt{2}$ states, Fig.\ref{f-1} (b).  From the mapping it is readily seen that $G_d (\omega)$ with $\sqrt{2}v_0=t$ and $\epsilon_0=0$ is identical to $G_0 (\omega)$ with $v_d=0$ when $U_0=U_d$.  However the conductances are different as shown in eqs.(\ref{eq:q0}) and (\ref{eq:embedded}).  The Kondo resonance in the embedded QD corresponds to the anti-Kondo resonance in the T-shaped QD at $\epsilon_0=0$.

It is easy to calculate the non-interacting case.  Since the Green's function is calculated as
\begin{eqnarray}
G_d (\omega)&=&{1\over \omega -\epsilon_d -v_d^2 g_0 (\omega)},
\end{eqnarray}
the conductance at $T=0$ is given by
\begin{eqnarray}
g&=&{1\over 1+q^2} {(e+q)^2 \over e^2 +1},
\label{eq:non-int:g}
\end{eqnarray}
with
\begin{eqnarray}
e&=& { \epsilon_d \over \Delta_d} - { \epsilon_0 \over \Delta_0}.
\end{eqnarray}

For interacting case at zero temperature we introduce the phase $e$ of the Green's function at $\omega=0$ as
\begin{eqnarray}
G_d(0)&=&{1\over \Delta_d} {1\over -e+\i},
\end{eqnarray}
where $\Delta_d$ is defined by eq.(\ref{eq:non-int:Delta_d}).  This is valid for any finite $U_d$, which is an essence of the Kondo effect.  The conductance for general $U_d$ at the zero temperature limit is given by the same formula, eq.(\ref{eq:non-int:g}).  These formulae of $g$ and $q$ are the same as the non-interacting case.  What is different from the non-interacting case is $e$ which contains all effects of interaction at zero temperature and defines the phase shift ${\pi \over 2} \Delta N$:
\begin{eqnarray}
e&=&\cot\left ({\pi \over 2} \Delta N\right),
\label{eq:phase-shift}
\end{eqnarray}
where $\Delta N$ is the total number of up and down electrons induced by the impurity, i.e. $0\le\Delta N\le2$.  The definition is $\Delta N=\langle N \rangle- \langle N \rangle_{\rm bulk}$, where $\langle N \rangle_{\rm bulk}$ is defined by the Hamiltonian without the $d$-site.  The present relation is nothing but the Friedel sum rule.

To study the conductance of the T-shaped QD numerically, we use the Finite Temperature Density Matrix Renormalization Group (F$T$-DMRG) method.  This method has an advantage that we can obtain results for different impurities in the same host one-dimensional system only by taking the expectation value concerning the impurity part.  Since a thermal Green's function can be calculated with the F$T$-DMRG, the local Green's function $G_d (\omega)$ can be obtained from the thermal Green's function $G_d (\tau)$ with numerical analytic continuation, where we use both the Pad{\'e} approximation and the Maximum Entropy Method (MEM).  For the F$T$-DMRG, the trotter number $M\le 200$ and the residual bases $m< 145$ are used.  For further details of this method we refer the readers to ref.~\citen{MaruyamaSU04}.

\begin{figure}[tb]
\resizebox{8cm}{!}{\includegraphics{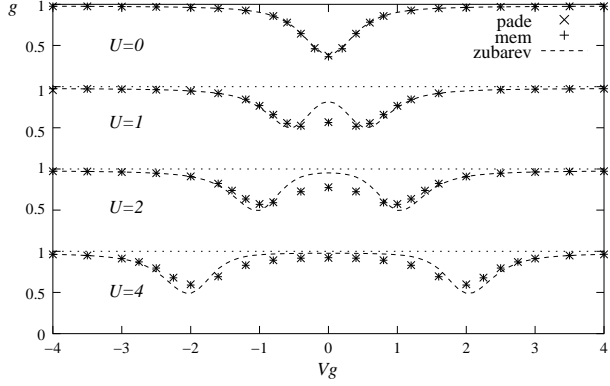}}
\caption{Conductance of T-shaped QD for $U_d=0,1,2,4$ as a function of gate voltage $V_g=\epsilon_d+U/2$ with $q=0$, $\Delta_0/t=1$ and $\Delta_d/t=0.25$ at $T=0.15$.
}
\label{f-2}
\end{figure}

\begin{figure}[tb]
\resizebox{8cm}{!}{\includegraphics{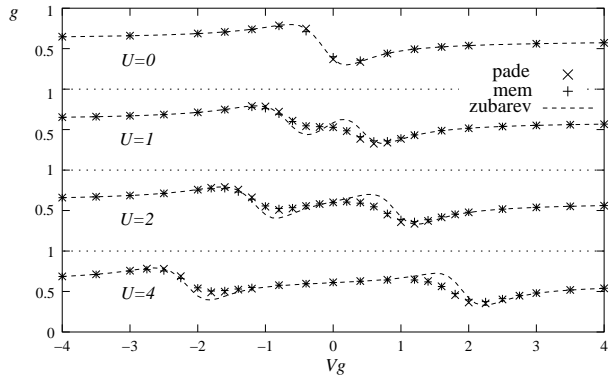}}
\caption{Conductance with $q=-0.8$  and $\Delta_d/t=0.15$.
Other conditions are the same as Fig.2}
\label{f-3}
\end{figure}

At high temperatures conductance obtained by the F$T$-DMRG shows typically two structures at $V_g=\pm U/2$, i.e. two dip structures for $q=0$ (Fig.~\ref{f-2}) and two Fano line shapes for $q\ne 0$ (Fig.~\ref{f-3}).  In these figures, we plot also conductance obtained by the Zubarev approximation:
\begin{eqnarray}
G_d (\omega)&=& \left (F (\omega) -v_d^2 g_0 (\omega)\right)^{-1} \\ F (\omega)&=&\left ( { 1- \langle n_d \rangle \over \omega -\epsilon_d} +{ \langle n_d \rangle \over \omega -\epsilon_d -U }\right)^{-1},
\label{eq:zubarev}
\end{eqnarray}
where $\langle n_d \rangle = \langle n_{d\uparrow} \rangle= \langle n_{d\downarrow} \rangle $ is the number of local electrons.  In the figures $\langle n_d\rangle$ obtained by the F$T$-DMRG is used in order to avoid further approximation to calculate $\langle n_d \rangle$.  These figures show perfect agreement at $U_d =0$ data between the F$T$-DMRG and the Zubarev's approximation which is exact in the noninteracting case.  The Zubarev's approximation shows reasonable agreement with the F$T$-DMRG results at larger $U_d$.  Characteristic features of conductance at high temperatures may be understood from the Zubarev's approximation.
However, it should be noted that the Zubarev's approximation is not a controlled scheme even at high temperatures, which is the origin of the differences between the F$T$-DMRG results and the Zubarev's approximation observed in Figs.\ref{f-2} and \ref{f-3}.

\begin{figure}[tb]
\resizebox{8cm}{!}{\includegraphics{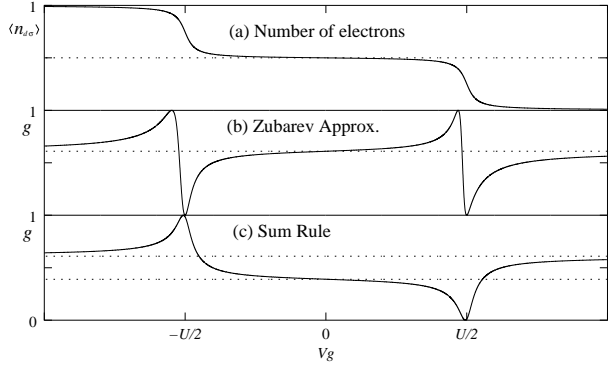}}
\caption{(a) Number of electrons as test function for $\langle n_d\rangle=\Delta N/2$.
(b) The conductance of the Zubarev's approximation with $q=-0.8$ and  $\langle n_d\rangle$ defined in (a),
which shows characteristic behaviours in the high temperature regime.
(c)  The conductance in the low temperature regime given by the sum rule from $\Delta N/2$ defined in (a).}
\label{f-4}
\end{figure}

Both the Zubarev's approximation, eq.(\ref{eq:zubarev}), and the zero temperature expression, eq.(\ref{eq:phase-shift}), are written by the number of electrons, $\langle n_d\rangle$ and $\Delta N$.  Let us suppose that temperature dependence of the number of electrons can be ignored and $\langle n_d\rangle$ is given by Fig.~\ref{f-4}(a).
At high temperatures we see two Fano structures at around $-U/2$ and $U/2$ on the background conductance of ${1\over 1+q^2}$, Fig.~\ref{f-4}(b).  At low temperatures we see development of a new Fano-Kondo plateau of ${q^2\over 1+q^2}$, Fig.~\ref{f-4}(c). At the same time the structure at $-U/2$ ($U/2$) becomes a simple peak (dip).

\begin{figure}[tb]
\resizebox{8cm}{!}{\includegraphics{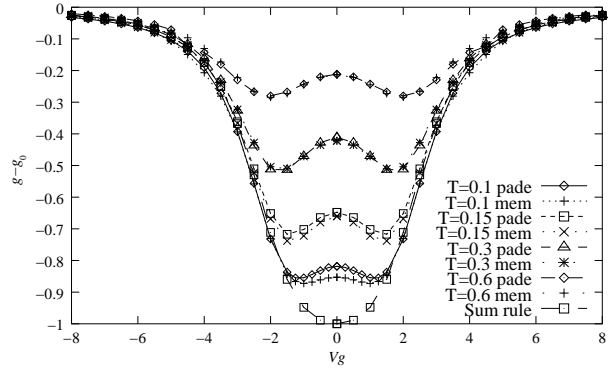}}
\caption{
Numerical results of conductance as a function of $V_g$ for $q=0$, $U_d=4$ and $\Delta_0/t=\Delta_d/t=1$ at various temperatures calculated by the F$T$-DMRG with the MEM and the Pad\'e approximation.
$g_0$ is the conductance of the background defined at $|V_g|=\infty$ limit.
}
\label{f-5}
\end{figure}

\begin{figure}[tb]
\resizebox{8cm}{!}{\includegraphics{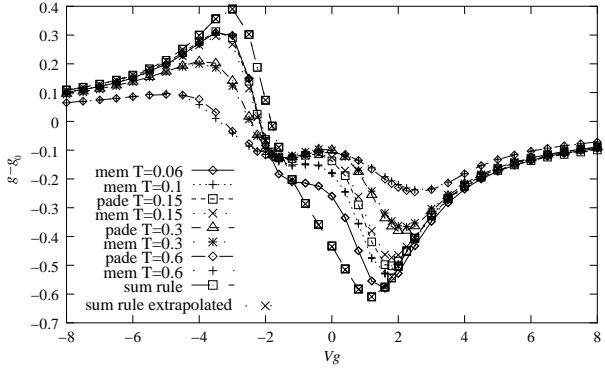}}
\caption{
Numerical results of conductance for $q=-0.8$, $U_d=4$, $\Delta_0/t=1$ and $\Delta_d/t=0.61$.  At low temperatures, $T=0.06$ and 0.1, only the results by the MEM are plotted due to the numerical instability of the Pad\'e approximation.  Other conditions are the same as Fig.5.
$g_0$ is the conductance of the background defined at $|V_g|=\infty$ limit.
}
\label{f-6}
\end{figure}

Numerical results of temperature dependence of conductance are plotted in Fig.~\ref{f-5} for $q=0$ and in Fig.\ref{f-6} for $q=-0.8$.
Differences between the results obtained by the Pad\'e approximation and the MEM may be considered as typical errors of the numerical calculations.  In Fig.~\ref{f-6}, at lower temperatures only the results by the MEM are shown due to the numerical instability of the Pad\'e approximation.
These figures show smooth crossover from high temperature regime to the zero temperature limit.  In these examples relatively large $\Delta_d$ is used in order to achieve the low temperature regime within the present numerical scheme.  Actually, the $q=0$ anti-Kondo resonance was recently observed by Katsumoto {\it et al.}~\cite{FanoQD:Katumoto} in a system with relatively large $\Delta_d/U$, the results of which are qualitatively consistent with Fig.\ref{f-5}.

At zero temperature, conductance is given by $\Delta N$ due to the Friedel sum rule, eq.(\ref{eq:phase-shift}).  $\Delta N$ is obtained from the numerical differentiation of the Free energy by changing the chemical potential $\mu$, $\langle N \rangle = - \delta F/\delta \mu$.  
Conductance obtained from $\Delta N$ evaluated at $T=0.02t$ is shown in Figs.\ref{f-5} and \ref{f-6}.
In Fig.\ref{f-6} the conductance obtained by $\Delta N$ extrapolated to $T=0$ is also shown.
The difference is so small and cannot be distinguished.  The Fano-Kondo plateau at ${q^2\over 1+q^2}$ is not clear for this set of parameters.  The main reason is relatively large $\Delta_d/U_d$.  Another reason is that for a finite $q$ the contribution from $\mbox{Re}[G_d]$ has a finite slope around $V_g=0$.  This effect becomes important at $q \sim \pm 1$.  We note that the center of this plateau where $\Delta N=1$ is shifted from $V_{g}=0$.

\begin{figure}[tb]
\resizebox{8cm}{!}{\includegraphics{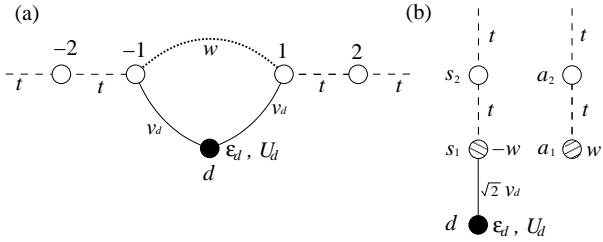}}
\caption{(a) The model for the Aharonov-Bohm quantum dot. (b) Mapping to two semi-infinite chains by the symmetric $s_i$ and antisymmetric $a_i$ states, $s_i=(c_i+c_{-i})/\sqrt{2}$ and $a_i=(c_i-c_{-i})/\sqrt{2}$.}
\label{f-7}
\end{figure}

Finally we comment on the AB-QD.  The simplest tight binding model for this case is shown in Fig.~\ref{f-7} (a).  For $T\ll t$ the transmission probability is given by eq.(\ref{eq:TlowT}) with
\begin{eqnarray}
q&=&{1-w^2 \over 2w}.
\end{eqnarray}
The model for the AB-QD may be mapped to two semi-infinite chains by introducing the symmetric, $s_i$, and the antisymmetric, $a_i$, sites.  From the mapping it is readily seen that $G_d (\omega)$ in the present model is essentially the same as the T-shaped QD with the identification
\begin{eqnarray}
\Delta_d&=& {2v_d^2 \Delta_0 \over \Delta_0^2 + w^2},
\end{eqnarray}
where $\Delta_0=t$ in the present model.  Therefore results discussed in the present paper are valid also for the AB-QD.

\section*{Acknowledgment}
We are grateful to Prof. S. Katsumoto for showing us their results prior to publication.

\end{document}